\documentclass[12pt]{article}
\usepackage{graphicx}
\newcommand{\dbox}{\,\raise2pt\hbox{\fbox{\rule{2.5pt}{0pt}\rule{0pt}{2.5pt}}}\,}
\newcommand{\qed}{\,\raise0pt\hbox{\mbox{\rule{6.5pt}{6.5pt}}}}
\newcommand{\bra}[1]{\mbox{$\langle #1 |$}}
\newcommand{\ket}[1]{\mbox{$| #1 \rangle$}}

\begin{document}
\setlength{\baselineskip}{7mm}

\begin{titlepage}
 \begin{normalsize}
  \begin{flushright}
        UT-Komaba/12-5\\
        June 2012
 \end{flushright}
 \end{normalsize}
 \begin{LARGE}
   \vspace{1cm}
   \begin{center}
     Closed string field theory in $a$-gauge\\
   \end{center}
 \end{LARGE}
  \vspace{5mm}
 \begin{center}
    Masako {\sc Asano} %\footnote{\tt }%
            \hspace{3mm}and\hspace{3mm}
    Mitsuhiro {\sc Kato}$^{\dagger}$ %\footnote{\tt }
\\
      \vspace{4mm}
        {\sl Faculty of Science and Technology}\\
        {\sl Seikei University}\\
        {\sl Musashino-shi, Tokyo 180-8633, Japan}\\
      \vspace{4mm}
        ${}^{\dagger}${\sl Institute of Physics} \\
        {\sl University of Tokyo, Komaba}\\
        {\sl Meguro-ku, Tokyo 153-8902, Japan}\\
      \vspace{1cm}

  ABSTRACT\par
 \end{center}
 \begin{quote}
  \begin{normalsize}
We show that $a$-gauge, a class of covariant gauges developed for bosonic open string field theory, is consistently applied to the closed string field theory. A covariantly gauge-fixed action of massless fields can be systematically derived from $a$-gauge-fixed action of string field theory.  

\end{normalsize}
 \end{quote}

\end{titlepage}
\vfil\eject

%%%%%%%%%%%
\section{Introduction and summary}
String field theory is a natural framework for the non-perturbative formulation of string theory. 
Indeed recent discovery of various classical solutions of string field equations, such as tachyon vacuum and various kinds of deformed vacua, 
has been revealing the fruitful structure of the open string field landscape\,\cite{Kostelecky:1989nt, Sen:1999nx, Sen:2000hx, Takahashi:2002ez, Schnabl:2005gv, Kiermaier:2007ba, Kiermaier:2010cf}.\footnote{As for the superstring, there has been a lot of trials toward its gauge-fixing e.g.~\cite{Preitschopf:fc,Arefeva:1989cp,Kroyter:2012ni,Torii:2012nj,Kroyter:2009rn,Kohriki:2011zz,Kohriki:2011pp,Kroyter:2009zi,Berkovits:2001im,Michishita:2004by} and the references therein.}
Nevertheless it is fair to say that the connection between string field theory and ordinary field theories has not been fully clarified yet. For example, since the interaction in open string field theory is cubic and non-local, it is not clear how non-perturbative properties of the low-lying component fields deviate from those of the ordinary field theories beyond low-energy perturbative correspondence. In order to make such problem clearer, it is an important subject to develop the covariant formulation of string field theory in a parallel way with ordinary field theory as much as possible.

As one of the useful tools for such a study, 
the present authors\,\cite{Asano:2006hk,Asano:2008iu} developed `$a$-gauge,' a one-parameter family of covariant gauges in the bosonic open string field theory, which naturally reduces to the corresponding covariant gauges in ordinary gauge theory including the Landau gauge as well as the Feynman gauge for the massless gauge mode.
Since the operator ${\cal O}_a$ used for defining the $a$-gauge commutes with the Virasoro operator $L_0$, this type of gauge condition does not mix the level of string modes. Thus $a$-gauge is also useful for the numerical study of string field theory by the level truncation method\,\cite{Asano:2006hm,Kishimoto:2009cz} as an alternative to the Siegel gauge.

In the present paper, we will extend our construction of $a$-gauge to the bosonic closed string field theory, which is expected to be convenient for studying the gravity and associated fields as a covariant gauge field theory derived from the closed string field theory. Besides massless modes, gauge fixing procedure itself and consistent truncation method of degrees of freedom are important for studying closed string field theory, since the theory is more complicated than open string case and its landscape is less known.

%%%
%To consider the gauge fixing problem in a systematic manner, 
We, however, only deal with the quadratic term of the theory in the present paper, because it has the similar form as for the open string field theory unlike the complicated interaction terms. 
For the full theory it is non-trivial to see whether the same gauge condition works as well. Even in the open string field theory case we know that the Gribov-like phenomenon could emerge in the level truncation study which is particularly apparent in Siegel gauge~\cite{Asano:2006hm}. Therefor we postpone the full-theory argument in the future work.

Fortunately, the BRST operator $Q$ has the structure parallel to the case of open string field theory on the restricted Hilbert space ${\cal H}'_c$ (see the subsequent section for its definition) for closed strings on which the string field theory is defined consistently\,\cite{Siegel:1984xd, Zwiebach:1992ie}.
Thus, if we can find the appropriate SU(1,1) structure on ${\cal H}'_c$ which was crucial in the previous study of open strings, the construction of $a$-gauge should be straightforward.

In fact, we show that we can identify the SU(1,1) algebra on ${\cal H}'_c$ from which we define $a$-gauge and construct the corresponding BRST invariant gauge fixed action.
By extracting the massless modes part from the $a$-gauge fixed action,  we see that  a one-parameter family of covariantly gauge-fixed action for  the weak graviton field $h_{\mu\nu}$ with dilaton $\phi$ and for the anti-symmetric tensor field $B_{\mu\nu}$ are automatically obtained.
The gauge-fixed action for $h_{\mu\nu}$ with $\phi$ constitutes of gauge invariant terms, gauge-fixing term (a generalized de~Donder gauge), and the (anti-)ghost terms given in (\ref{eq:Sgf+1}), plus auxiliary fields terms (\ref{eq:Sgf+2}) which are decoupled from the other terms if we only consider the quadratic part of the action.
The action for $B_{\mu\nu}$ has the similar structure as given in (\ref{eq:Sgf-1}) and (\ref{eq:Sgf-2}), except that in this case there also appear the second (anti-)ghost fields which is necessary to fix the gauge invariance of the first (anti-)ghost term.
Note that the resulting massless part of the gauge invariant or gauge-fixed action can be applied in the spacetime of any dimension $D$ although the original string field theory action is consistent only for $D=26$.

%%%%%%%
\section{Structure of the BRST cohomology}
%%%%%%%
We first review the properties of Hilbert space of closed string field theory based on
ref.\cite{Zwiebach:1992ie, Zwiebach:1993cs}.
To construct the consistent Hilbert space for closed string field theory, 
we provide the direct product of two sets of open string state space as
\begin{equation}
{\cal H}_{\rm c}= {\cal H} \otimes \bar{\cal H} 
\end{equation}
and restrict ${\cal H}_{\rm c}$ to the space ${\cal H}'_{\rm c}$
of states $\ket{f_c} \in {\cal H}_{\rm c}$
which satisfy the conditions 
\begin{equation}
(b_0-\bar{b}_0)\ket{f_c}=0, \qquad (L_0-\bar{L}_0)\ket{f_c}=0.
\label{eq:condbL}
\end{equation}
Here, each state in ${\cal H}$ consists of   
$\alpha_{-l}^{\mu}$ $(l\ge 1)$, $c_{-m}$ $(m\ge -1)$, and $b_{-n}$ $(n\ge 2)$ operated on  
the state 
$\ket{0, p}=(e^{ip_\mu x^\mu} \ket{0}_{\rm m})\otimes (\ket{0}_{\rm g})$.
The states $\ket{0}_{\rm m}$ and $\ket{0}_{\rm g}$ denote the matter and the ghost ground states respectively.
The space $\bar{\cal H}$ 
consists of matter and ghost fields with bars 
($\bar{\alpha}_{-l}^{\mu}$, $\bar{c}_{-m}$ and $\bar{b}_{-n}$)
and has the same structure as ${\cal H}$.
Also, $L_0$ and $\bar{L}_0$ are given by 
\begin{equation}
L_0=\frac{\alpha'}{4}p^2+N-1, \qquad \bar{L}_0=\frac{\alpha'}{4}p^2+\bar{N}-1
\qquad\quad \left(\alpha_0^\mu=\sqrt{\frac{\alpha'}{2}}p^\mu\right)
\end{equation}
where $N$ (or $\bar{N}$) is the level of a state in ${\cal H}$ (or in $\bar{\cal H}$) as usual.

Non-degenerate inner product in the restricted space ${\cal H}'_{\rm c}$ is given by
\begin{equation}
\left\langle \ket{f}  , \ket{g} \right\rangle \equiv i \left\langle {\rm bpz}(f) | c_0^- g \right\rangle
\label{eq:clinnpro} 
\end{equation}
with $c_0^{\pm} \equiv \frac{1}{2}(c_0 \pm \bar{c}_0)$
and the relation 
\begin{equation}
 \langle 0,p| c_{-1}\bar{c}_{-1} c_0^+ c_0^- c_1\bar{c}_1 \ket{0,p'} =-\frac{i}{2}(2\pi)^D\delta^{(D)}(p-p')
\label{eq:innerpro}
\end{equation}
with $D=26$. Here, $\langle\mbox{bpz}(f)|$ $\equiv$ bpz($\ket{f}$) denotes the BPZ conjugate of $\ket{f}$ and is related to the Hermitian conjugate $\bra{f} $ by bpz($\ket{f}$)$=\epsilon_f\bra{f}$ with $\epsilon_f=\pm 1$ (See the Appendix C of \cite{Asano:2006hk}.) 
The $i$ factor in the right-hand side of eq.(\ref{eq:innerpro}) is necessary because the Hermitian conjugate of the left-hand side coincides with minus of itself:
$\langle 0| c_{-1}\bar{c}_{-1} c_0^+ c_0^- c_1\bar{c}_1 \ket{0}^* = - \langle 0| c_{-1}\bar{c}_{-1} c_0^+ c_0^- c_1\bar{c}_1 \ket{0} $. 
From the conditions given by eqs.(\ref{eq:condbL}), the space ${\cal H}'_{\rm c}$
is divided into two parts which are determined by whether 
$\tilde{c}_0 \equiv c_0+\bar{c}_0(=2c_0^+)$ is present or not,
{\it i.e.}~,
\begin{equation}
{\cal H}'_{\rm c} = \tilde{\cal F}_{\rm c} + \tilde{c}_0 \tilde{\cal F}_{\rm c}
\end{equation}
where $\tilde{\cal F}_{\rm c}$ is the set of states of the form 
\begin{equation}
\ket{f} = \ket{f_{\rm L}}  \otimes \ket{f_{\rm R}} 
\;\in\; b_0c_0{\cal H} \otimes \bar{b}_0\bar{c}_0\bar{\cal H}
,\qquad
N \ket{f}=\bar{N}\ket{f}. 
\label{eq:restf}
\end{equation}

In order to define the $a$-gauge, we divide the BRST operator $Q$
by the ghost zero modes as in the case of open string field theory \cite{Asano:2006hk}
\begin{equation}
Q=(\tilde{Q}+b_0 M +c_0L_0) + (\tilde{\bar{Q}}+\bar{b}_0 \bar{M} +\bar{c}_0\bar{L}_0).
\end{equation}
Here, $M$ and $\tilde{Q}$ (and the corresponding operators with bars)
are independent of ghost zero modes and explicitly given by 
\begin{equation} 
M=-2\sum_{n=1}^{\infty} n c_{-n} c_n,
\quad
\tilde{Q}= \sum_{n\ne 0} c_{-n} 
L_n^{({\rm m})}  -\frac{1}{2} \sum_{\parbox{12mm}{\tiny\rule{5pt}{0pt}$mn\ne 0$\\$m+n\ne0$}} (m-n)\,:c_{-m} c_{-n} b_{n+m}: .
\end{equation}
In the restricted state space ${\cal H}'_{\rm c}$,
$Q$ is further contracted to 
\begin{equation}
Q=\tilde{Q}^+ + \tilde{c}_0 L_0 + b_0 {\cal M}
\qquad (\mbox{on ~${\cal H}'_{\rm c}$})
\label{eq:QinH'}
\end{equation}
where we have used the notations
$\tilde{Q}^+ \equiv  \tilde{Q}+ \tilde{\bar{Q}}$ and
${\cal M} \equiv  M+ \bar{M}$. 
In the right-hand side of eq.(\ref{eq:QinH'}), $b_0$ and $L_0$ can be replaced by $\bar{b}_0$ and $\bar{L}_0$ respectively.
Note that the relation 
\begin{equation}
(\tilde{Q}^+)^2=-L_0 {\cal M} \quad (= -\bar{L}_0 {\cal M}  )
\end{equation}
holds on ${\cal H}'_c$.
Furthermore, in the space $\tilde{\cal F}_{\rm c} $ (or in $\tilde{c}_0 \tilde{\cal F}_{\rm c}$), 
operators
\begin{equation}
{\cal M}= -\sum_{n=1}^{\infty} 2n (c_{-n} c_n +\bar{c}_{-n} \bar{c}_n),
\end{equation}
\begin{equation}
{\cal M}^-\equiv -\sum_{n=1}^{\infty} \frac{1}{2n} (b_{-n} b_n +\bar{b}_{-n} \bar{b}_n)
\end{equation}
and 
\begin{equation}
{\cal M}_z\equiv 
\sum_{n=1}^{\infty} \frac{1}{2}(c_{-n} b_n -b_{-n} c_n 
+\bar{c}_{-n} \bar{b}_n -\bar{b}_{-n} \bar{c}_n)\;\;
\left( =  \frac{1}{2} \hat{G} \right)
\end{equation}
constitute the SU(1,1) algebra specified by 
\begin{equation}
 [{\cal M},{\cal M}^-] = 2 {\cal M}_z ,
\quad [{\cal M}_z,{\cal M}] = {\cal M}, 
\quad [{\cal M}_z,{\cal M}^-] = -{\cal M}^-.  
\end{equation}
The operator  $\hat{G}$ counts the ghost number of the non-zero mode part 
and the relation to the total ghost number operator $G$ (with $G\ket{0}=0$)
is given by $G= \hat{G}+c_0b_0+\bar{c}_0\bar{b}_0$.
We divide the space ${\cal F}_c$ by $\hat{G}$ as
\begin{equation}
{\cal F}_c =\bigoplus_{n=-\infty}^\infty {\cal F}_c^{\hat{G}=n}.
\end{equation}
Every state in ${\cal F}_c$ is decomposed into finite-dimensional irreducible representations of SU(1,1) characterized by a spin $s$ which is an integer or a half-integer.
A spin $s$ representation consists of $2s+1$ states each of which belongs to 
${\cal F}_c^{n}$ with $n=-2s, -2s+2,\cdots, 2s$.
Thus, as in the open string case, all the states in ${\cal F}_c^{n}$ belong to either of the spin $s= \frac{|n|}{2}+ i$ ($i\in \mbox{\boldmath Z}$, $i\ge 0$) representation, and 
there is an isomorphism between two spaces ${\cal F}_c^{n}$ and  ${\cal F}_c^{-n}$ $(n>0)$.
This is described by using the operator ${\cal W}_n$ which is defined to be 
`inverse'  of ${\cal M}^n$ as 
\begin{equation}
 {\cal M}^n {\cal W}_n {\cal F}_c^{n} ={\cal F}_c^{n},
\qquad 
{\cal W}_n {\cal M}^n {\cal F}_c^{-n} ={\cal F}_c^{-n}.
\end{equation}
For a state $ \ket{f^n}_s \in {\cal F}_c^n $ with a definite spin $s$ $(\ge n/2)$, 
the operation of ${\cal W}_n$ is nothing but $(M^-)^n$ times a constant factor depending on $s$, {\it i.e.}, ${\cal W}_n  \ket{f^n}_s \propto (M^-)^n \ket{f^n}_s $.
For a general state which might be a sum of the states with different spins, 
${\cal W}_n$ is represented by using ${\cal M}$ and ${\cal M}^-$ as
\begin{equation}
 {\cal W}_n = 
\sum_{i=0}^{\infty}\, (-1)^i \frac{(n+i-1)!}{[(n+i)!]^2\, i!\,(n-1)!}\,
({\cal M}^-)^{n+i} {\cal M}^i .
\end{equation}

From the above discussion, we see that 
the BRST cohomology on the restricted state space ${\cal H}_c'$ has 
exactly the same property as for the open string field theory 
in that for both cases the BRST operator $Q$ and the state space 
can be divided with respect to two ghost zero modes: 
$b_0$ (or $\bar{b}_0$) and $\tilde{c}_0$ for the closed string theory whereas 
$b_0$ and $c_0$ for the open string theory.
With this property, we can construct $a$-gauge for the closed string field theory in a way parallel to the open string field theory case\,\cite{Asano:2006hk}.

%%%%%%%
\section{The $a$-gauge and the gauge fixed action}
%%%%%%%
The consistent gauge invariant action (quadratic part) for closed string theory is given by
\begin{equation}
S=\frac{1}{2}\langle \Phi_2 , Q\Phi_2 \rangle 
\label{eq:action2}
\end{equation}
if we restrict the string field $\Phi_2 \in {\cal H}'_c$ and use the inner product defined by eq.(\ref{eq:clinnpro}).
More precisely, 
$\Phi_2 $ is expanded by the ghost number $G=2$ states as 
\begin{equation}
\Phi_2 = \int \frac{d^{D}p }{(2\pi)^D}\sum_i \ket{f^i} \phi_{f^i}(p),
\qquad  \ket{f^i} \in {\cal H}'_c (p).
\label{eq:expandstatefield2}
\end{equation}
The Grassmann parity of a field $\phi_{f^i}(p)$ is determined to be the same as that of $\ket{f^i}$ so that $\Phi_2$ becomes Grassmann even.
In order that the action $S$ is hermitian and has the correct sign factor, we should assign  the appropriate hermiticity for each field $\phi_{f^i}(p)$:
For a state $\ket{f^i}$ which has the property bpz($\ket{f}$)$=\epsilon_f \bra{f}$, the corresponding field $\phi_{f}(p)$ should have the property 
$(\phi_{f}(-p))^* =  - \epsilon_f \phi_{f}(p)$. 

The action eq.(\ref{eq:action2}) is invariant under the gauge transformation
\begin{equation}
\delta \Phi_2 = Q \Lambda_1
\label{eq:gaugetr}
\end{equation}
for any $G=1$ Grassmann odd string field $\Lambda_1$.
This gauge invariance is precisely fixed if we impose the condition 
\begin{equation}
%\frac{1}{1-a}(b_0 +ab_0\tilde{c}_0 {\cal W}_{n-1}{\cal M}^{n-2}\tilde{Q}^+) \Phi_2 =0
\frac{1}{1-a}(b_0 +a\,b_0\tilde{c}_0 {\cal W}_{1}\tilde{Q}^+) \Phi_2 =0
\end{equation}
with any value of the parameter $a$ (including $a=\infty$) except $a=1$ at which the combination of operators appeared in the above equation becomes gauge invariant and it does not give a gauge fixing condition.
The validity of this condition as a gauge fixing condition is proved
similarly as in the case of open string field theory\,\cite{Asano:2006hk}
since ${\cal H}_c'$ has the structure parallel to the open string state space as we saw in the previous section.
In particular, for $a=0$, it reduces to the well-known Siegel gauge condition.

We construct the gauge fixed action corresponding to the above condition for general $a$  
by the similar scheme used for the open string field theory case\,\cite{Asano:2006hk}.
We first provide the extended action by including the string fields with all the ghost numbers as
\begin{equation}
\tilde{S} =
{1 \over 2}  \sum_{n=-\infty}^{\infty} \left\langle \Phi_n, Q  \Phi_{-n+4} \right\rangle
\end{equation}
which is invariant under the extended gauge transformation
\begin{equation}
\delta \Phi_n = Q \Lambda_{n-1}.
\label{eq:gaugetr_ex}
\end{equation}
This extended gauge transformation is appropriately fixed by imposing the condition
\begin{equation}
{\rm bpz}\left( {\cal O}_a^{\langle 6-n \rangle}\right) \Phi_n =0
\label{eq:gaugecondg}
\end{equation}
where ${\cal O}_a^{\langle n \rangle}$ is given by
\begin{equation}
{\cal O}_a^{\langle n+2 \rangle} = 
\frac{1}{1-a}(1 +a\, \tilde{Q}^+   {\cal M}^{n-2} {\cal W}_{n-1} \tilde{c}_0)b_0 \hspace*{2cm} (n\ge 2)
\end{equation}
and
\begin{eqnarray}
{\cal O}_{|a|<\infty}^{\langle -n+5 \rangle}  &=& 
(1  +a\, {\cal W}_{n-1} {\cal M}^{n-2} \tilde{Q}^+ \tilde{c}_0 ) b_0 
\hspace*{2cm} (n\ge 2),
\label{eq:O-afinite}
\\
{\cal O}_{a=\infty}^{\langle -n+5 \rangle}  &=& 
(1-{\cal P}_{-n+2}) b_0  - a\, {\cal W}_{n-1} {\cal M}^{n-2} \tilde{Q}^+ \tilde{c}_0  b_0 
\qquad (n\ge 2).
\end{eqnarray}
In the above equations, ${\cal P}_{-n}$ ($n\ge 0$) is the projection operator 
\begin{equation}
{\cal P}_{-n}= \frac{1}{L_0} \tilde{Q}^+ {\cal W}_{n+1} \tilde{Q}^+ {\cal M}^{n} 
\end{equation}
on the space $\tilde{\cal F}_c^{\hat{G}=-n}$ ($n\ge 0$).
We chose the form of ${\cal O}_{|a|<\infty}^{\langle -n+5 \rangle}$ ($n\ge 2$) in eq.(\ref{eq:O-afinite}) 
as given in ref.\cite{Asano:2006hk} rather than the form used in ref.\cite{Asano:2008iu}.
This is to avoid unnecessary non-local field redefinitions when rewriting the 
gauge-fixed action in a simpler form.
The gauge fixed action corresponding to the condition (\ref{eq:gaugecondg}) is then given by 
\begin{equation}
S^a_{\rm GF} =
{1 \over 2}  \sum_{n=-\infty}^{\infty} \left\langle \Phi_n, Q  \Phi_{-n+4} \right\rangle
-  \sum_{n=-\infty}^{\infty} 
\left\langle {\cal O}_a^{\langle -n+6 \rangle}  {\cal B}_{-n+6} , \Phi_{n}  \right\rangle 
\label{eq:gfaction}
\end{equation}
where ${\cal B}_{n}$ are the $G=n$ Grassmann even string fields which satisfy ${\cal B}_{n} = \tilde{c}_0b_0 {\cal B}_n$.
Note that each of the string fields $\Phi_n$ and ${\cal B}_{n}$  is expanded by string states with $G=n$ as in eq.(\ref{eq:expandstatefield2}).
The hermiticity of a field $\phi_{f}(p)$ in $\Phi_n$ 
is determined in the same way as for $\Phi_2$:
For a field $\phi_{f}(p)$ associated with a state $\ket{f}$,
we assign $(\phi_{f}(-p))^* =  - \epsilon_f \phi_{f}(p)$ if bpz($\ket{f}$)$=\epsilon_f \bra{f}$. 
On the other hand, for a field ${\beta_f}(p)$ in ${\cal B}_n$ which is associated with 
a state $\ket{f}$, 
we assign $({\beta_f}(-p))^* =  + \epsilon_f {\beta_f}(p)$
when bpz($\ket{f}$)$=\epsilon_f \bra{f}$.

As in the open string case\,\cite{Asano:2006hk,Asano:2008iu},
the gauge fixed action $S^a_{\rm GF}$ is invariant under the BRST transformations given by
\begin{equation}
\delta_B \Phi_n = \eta {\cal O}_a^{\langle n+1 \rangle}  {\cal B}_{n+1} \; (n>2),
\qquad
\delta_B \Phi_{n} = \eta Q \Phi_{n-1} \; (n\le 2),
\qquad
\delta_B {\cal B}_n = 0
\label{eq:BRS1}
\end{equation}
with a Grassmann odd parameter $\eta$.

Some comments on the validity of the above gauge fixing are in order.
Firstly, any string field $\Phi = \sum_n \Phi_n$ with extended ghost number can be uniquely transformed to satisfy the gauge condition (\ref{eq:gaugecondg}) by the gauge transformation (\ref{eq:gaugetr_ex}) provided $L_0\ne 0$, and there is no residual gauge degrees of freedom remained. Secondly, all extra fields other than physical degrees of freedom contained in $\Phi_2$ are either directly killed by the gauge condition (\ref{eq:gaugecondg}) or become a member of the BRST quartets for the BRST transformation (\ref{eq:BRS1}). This structure is common for all the value of $a$ including Siegel gauge ($a=0$).

%%%%%%%%%%%%%%%%%%%%%

%%%%%%%%%%%%%%%%%%%%%%%%%%%%
\section{Example: level one part of the action}
We will explicitly see the level $N=1$ massless part of the action. 
We first consider the gauge invariant action (\ref{eq:action2}). 
The $N=1$ part of the string field $\Phi_2^{N=1}$ with $G=2$ is expressed as the sum of the two parts as 
\begin{equation}
\Phi_2^{N=1}= \Phi_{2,+}^{N=1} + \Phi_{2,-}^{N=1} 
\label{eq:N2Phipm}
\end{equation}
where
\begin{eqnarray}
\Phi_{2,+}^{N=1} &=&\int\!\frac{d^D p}{(2\pi)^D} 
\left[
\frac{1}{\sqrt{\alpha'}}
\alpha_{-1}^\mu\bar{\alpha}_{-1}^\nu\ket{p;\downarrow\downarrow}h_{\mu\nu}(p)
+
\frac{1}{\sqrt{\alpha'}}
(b_{-1}\bar{c}_{-1} -c_{-1}\bar{b}_{-1})\ket{p;\downarrow\downarrow}\phi_+(p)
\right.
\nonumber\\
&& \qquad+
\left.
\frac{i}{2}\tilde{c}_0^+ (\alpha_{-1}^\mu\bar{b}_{-1} +b_{-1}\bar{\alpha}^\mu_{-1}) 
\ket{p;\downarrow\downarrow}\omega_{\mu}^+(p)
\right]
\label{eq:N1Phi+}
\end{eqnarray}
and
\begin{eqnarray}
\Phi_{2,-}^{N=1} &=&\int\!\frac{d^D p}{(2\pi)^D} 
\left[
\frac{1}{\sqrt{\alpha'}}
\alpha_{-1}^\mu\bar{\alpha}_{-1}^\nu\ket{p;\downarrow\downarrow}B_{\mu\nu}(p)
+
\frac{1}{\sqrt{\alpha'}}
(b_{-1}\bar{c}_{-1} +c_{-1}\bar{b}_{-1})\ket{p;\downarrow\downarrow}\phi_-(p)
\right.
\nonumber\\
&&\qquad +
\left.
\frac{i}{2}\tilde{c}_0^+(\alpha_{-1}^\mu\bar{b}_{-1} -b_{-1}\bar{\alpha}^\mu_{-1}) 
\ket{p;\downarrow\downarrow}\omega_{\mu}^-(p)
\right].
\label{eq:N1Phi-}
\end{eqnarray}
In the above equations, 
$h_{\mu\nu}=h_{\nu\mu}$, $B_{\mu\nu}=-B_{\nu\mu}$
and $\ket{p;\downarrow\downarrow}= c_1\bar{c}_1\ket{p; 0}$.
We have divided the string fields $\Phi_{2}^{N=1}$ into ``$+$'' and ``$-$''-parts as above in order that the gauge invariant action as well as the gauge transformation are divided into two independent parts.
In fact, the action (\ref{eq:action2}) for $\Phi_+^{N=1}$ and $\Phi_-^{N=2}$ parts are decoupled from each other and can be written as
\begin{equation}
S_{N=1}= \frac{1}{2} \langle \Phi_{2,+}^{N=1}  , Q \Phi_{2,+}^{N=1} \rangle
+ \frac{1}{2}\langle \Phi_{2,-}^{N=1}  , Q \Phi_{2,-}^{N=1} \rangle
\quad \left(\equiv S_{N=1}^+ + S^-_{N=1} \right)
\end{equation}
and, as we will see, each action $S_{N=1}^+$ or $S^-_{N=1} $ is invariant under independent gauge transformation.

The ``$+$''-part of the gauge invariant action $S_{N=1}^+$ is explicitly given in the position space representation by
\begin{eqnarray}
S_{N=1}^+ &=&
\frac{1}{2}\int d^Dx \,e^{ipx} \langle \Phi_{2,+}^{N=1}  , Q \Phi_{2,+}^{N=1} \rangle
\nonumber\\
& = &
\frac{1}{2}\int d^Dx \left[
\left[\sqrt{-g'}R'\right]\!\Bigg|_2 
-\frac{1}{D-2}\partial_\mu \phi_+' \partial^\mu \phi'_+ 
  - \omega'^{+}_{\mu}\omega'^{+\mu}
\right]
\label{eq:action+inv} 
\end{eqnarray}
where we have defined 
\begin{equation}
\phi_+' = \phi_+ + \frac{1}{2}h_\mu{}^{\mu} 
,\qquad
h'_{\mu\nu} = h_{\mu\nu} -\frac{1}{D-2} \eta_{\mu\nu} \phi_+
,\qquad
\omega_{\mu}'^{+} =\omega_{\mu}^{+} 
  + \frac{1}{\sqrt{2}} \partial^{\nu}(h_{\mu\nu}+\eta_{\mu\nu} \phi_+)
\end{equation}
from the original fields given in eq.(\ref{eq:N1Phi+}).
In eq.(\ref{eq:action+inv}), $\left.[\sqrt{-g'}R']\right|_2$is the 
Einstein action term with metric $g'_{\mu\nu}\sim \eta_{\mu\nu}+h'_{\mu\nu}$
expanded up to the quadratic order with respect to $h'$.
The gauge transformation eq.(\ref{eq:gaugetr}) for $S_{N=1}^+$ is reduced to 
\begin{equation}
\delta h'_{\mu\nu} = 2 \partial_{(\mu}\lambda^+_{\nu)}.
\end{equation}
Note that $\omega_{\mu}'^{+}$ and $\phi'_+$ are gauge invariant.
On the other hand, 
the ``$-$''-part of the action $S_{N=1}^-$ is given by
\begin{eqnarray}
S_{N=1}^- &=&
\frac{1}{2}\int d^Dx \,e^{ipx} \langle \Phi_{-}^{N=1}  , Q \Phi_{-}^{N=1} \rangle
\nonumber\\
& = &
\int d^Dx \left[
-\frac{1}{24} H_{\mu\nu\rho} H^{\mu\nu\rho} 
-\frac{1}{2} \omega_{\mu}'^{-}\omega'^{-\mu}
\right]
\label{eq:action-nv} 
\end{eqnarray}
where 
\begin{equation}
H_{\mu\nu\rho}=3\partial_{[\mu}B_{\nu\rho]},
\qquad
\omega_{\mu}'^{-} =\omega_{\mu}^{-} 
  + \frac{1}{\sqrt{2}} (\partial^\nu B_{\mu\nu}+\partial_\mu \phi_-).
\end{equation}
In this case, the gauge transformation is given by
\begin{equation}
\delta B_{\mu\nu} = 2\partial_{[\mu}\lambda_{\nu]}^- .
\end{equation}
Note that $H_{\mu\nu\rho}$ and $\omega_{\mu}'^{-}$ are the gauge invariant combinations
and the scalar field $\phi_-$ only appears in the action eq.(\ref{eq:action-nv}) via $\omega_\mu'^{-}$.

%%%%%
Next, we consider the gauge fixed action $S_{{\rm GF},N=1}^a $,
which can be read from 
eq.(\ref{eq:gfaction}) by expanding the $N=1$ string fields $\Phi_n^{N=1}$
with respect to the component fields.
%with respect to $N=1$ string states with the corresponding fields.
% and calculating the inner products of the string states.
For this level $N=1$, non-trivial string fields are given by $\Phi_{n}$ for $0\le n \le 5$ and ${\cal B}_n$ for $1\le n \le 5$.
As for the string field $\Phi_2$ in eq.(\ref{eq:N2Phipm}), we can divide all these non-trivial 
string fields into two parts as $\Phi_n^{N=1}= \Phi_{n,+}^{N=1} + \Phi_{n,-}^{N=1} $
and ${\cal B}_n = {\cal B}_{n,+}^{N=1} + {\cal B}_{n,-}^{N=1} $ so that 
the ``$+$'' and the ``$-$''- parts of the fields only contribute to the gauge fixed actions for $S^{N=1}_{2,+}$ and $S^{N=1}_{2,-}$, respectively.

The resulting action for the ``$+$"-part is given by
\begin{eqnarray}
S_{{\rm GF}, N=1,+}^{a} &=&
 \int d^Dx \left[
\frac{1}{2} 
\left[\sqrt{-g'}R'\right]\!\Bigg|_2 
-\frac{1}{2(D-2)}\partial_\mu \phi'_+ \partial^\mu \phi'_+ 
+\left(\partial^\nu h'_{\mu\nu} -\frac{1}{2}\partial_\mu h_{\nu}'^{\nu}\right) \beta^{+\mu}
\right.
\nonumber\\
&  & \qquad\qquad
\left.
+\frac{1}{(1-a)^2} \beta_{\mu}^+ \beta^{+\mu}
+\frac{i}{2} \partial_\nu\bar{\gamma}_{\mu}'^{+} \partial^\nu {\gamma}'^{+\mu}
\right]
\label{eq:Sgf+1}
\\
&  &
+ \int d^Dx \left[
  - \frac{1}{2} \omega''^{+}_{\mu}\omega''^{+\mu}
-i\bar{\omega}_{(\mu\nu)}^{\rm o} {\beta}^{{\rm o}(\mu\nu)} 
-i \bar{\omega}_+^{\rm o} {\beta}_+^{\rm o} 
-\frac{1}{2}\bar{g}_{\mu}^ +\beta_{g}^{+\mu}
\right].
\label{eq:Sgf+2}
\end{eqnarray}
The first part (\ref{eq:Sgf+1}) is nothing but a generalization of the de~Donder (or harmonic) gauge action of the graviton.%%%
\footnote{$a=0$ corresponds to what is called de~Donder gauge in many literatures, although the de Donder's gauge condition was originally a classical notion defined as a gravitational counterpart of Lorentz gauge condition in the classical electromagnetism.}
Here, the vector field $\beta_{\mu}^+$ which belongs to ${\cal B}_{4,+}$ is
the auxiliary field for $h_{\mu\nu}'$ and is 
corresponding to the Nakanishi-Lautrup field for the vector gauge theory.
By using this $\beta_{\mu}^+$, we have defined 
\begin{equation}
\omega_{\mu}''^{+} =\omega_{\mu}'^{+} 
+\frac{\sqrt{2}}{1-a}\beta_{\mu}^+.
\end{equation}
The Grassmann odd fields ${\gamma}_{\mu}'^{+}$ and $\bar{\gamma}_{\mu}'^{+}$, which respectively belong to $b_0\tilde{c}_0\Phi_{1,+}$ and  $b_0\tilde{c}_0\Phi_{3,+}$, 
are the ghost and the anti-ghost fields for the gauge invariance of $h_{\mu\nu}'$.
The other fields are from $\tilde{c}_0b_0\Phi_{n,+}$ and ${\cal B}_{6-n,+}$ with $n=3,\,4$.
The fields in $\tilde{c}_0b_0\Phi_{3,+}$ and ${\cal B}_{3,+}$  are collected as   
the Grassmann odd symmetric tensor fields $\bar{\omega}_{(\mu\nu)}^{\rm o}$ and ${\beta}^{{\rm o}}_{(\mu\nu)}$, and the Grassmann odd scalar fields $\bar{\omega}^{\rm o}$ and ${\beta}^{\rm o}$ after suitable field redefinition.  
For $n=4$, there are two Grassmann even vector fields: 
$\bar{g}_{\mu}^ +$ from $\tilde{c}_0b_0\Phi_{4,+}$ and $\beta_{g}^{+\mu}$ from ${\cal B}_{2,+}$.
In the action $S_{{\rm GF}, N=1,+}^{a}$,
the first two lines (\ref{eq:Sgf+1}) form the minimal gauge fixed action for 
the original gauge invariant action eq.(\ref{eq:action+inv}).
On the other hand, the terms in the third line (\ref{eq:Sgf+2}) only contain auxiliary fields and they decouple from the main part (\ref{eq:Sgf+1}).

%%%%%%
The ``$-$"-part of the gauge fixed action is given by%
\footnote{Note that the $a$ in the second line of eq.(\ref{eq:Sgf-1}) 
can be replaced by arbitrary value $a'$ since the gauge condition can be taken  independently for the fields associated with different SU(1,1) spin string modes.
Such an extension of the $a$-gauge is given in the forthcoming paper\,\cite{ma} by one of the present authors.}
\begin{eqnarray}
S_{{\rm GF}, N=1,-}^{a} &=& 
\int d^Dx \left[
-\frac{1}{24} H_{\mu\nu\rho} H^{\mu\nu\rho} 
-(\partial^{\mu}B_{[\mu\nu]}-\partial_\nu \phi^-)\beta_-^\nu
+\frac{1}{(1-a)^2} \beta_{\mu}^- \beta^{-\mu}
\right.
\nonumber\\
&&
+i \partial_{[\mu} \bar{\gamma}_{\nu]}^{-} \partial^{[\mu} {\gamma}^{-\nu]}
-\frac{i}{2} \beta^{{\rm o}-} \partial_{\mu} \bar{\gamma}^{-\mu }
-\frac{i}{2} \bar{\beta}^{{\rm o}-} \partial_{\mu} {\gamma}^{-\mu}
+\frac{i}{1-a} \bar{\beta}^{{\rm o}-} {\beta}^{{\rm o}-}
\nonumber\\
&&
\left.
-\frac{1}{2} \partial_{\mu} \bar{f} \partial^{\mu} {f} 
\right]
\label{eq:Sgf-1}
\\
&&
+
\int d^Dx \left[
-\frac{1}{2} \omega_{\mu}''^{-}\omega''^{-\mu}
-i\bar{\omega}^{\rm o}_{[\mu\nu]}{\beta}^{{\rm o}[\mu\nu]} 
-i \bar{\omega}^{{\rm o}-} {\omega}^{{\rm o}-} 
-\frac{1}{2}\bar{g}_{\mu}^- \beta_{g}^{-\mu}
+i\bar{\theta} \beta_{\theta}
\right].
\label{eq:Sgf-2}
\end{eqnarray}
In this equation, the fields $B_{[\mu\nu]}$, $\phi^-$ and $\omega_{\mu}'^{-}$ from $\Phi_{2,-}$ with the field $\beta_\mu^-$ from ${\cal B}_{4,-}$ are collected in the first three terms of (\ref{eq:Sgf-1}) and the first term of (\ref{eq:Sgf-2}).
As for the ``$+$"-part action, the field $\beta_\mu^-$ corresponds to the Nakanishi-Lautrup field and we have defined $\omega_{\mu}''^{-}$ as
\begin{equation}
\omega_{\mu}''^{-} =\omega_{\mu}'^{-} +\frac{\sqrt{2}}{1-a} \beta_{\mu}^-.
\end{equation}
We also have the Grassmann odd ghost and the anti-ghost fields 
${\gamma}_{\mu}'^{-}$ and $\bar{\gamma}_{\mu}'^{-}$ which are from $b_0\tilde{c}_0\Phi_{1,-}$ and $b_0\tilde{c}_0\Phi_{3,-}$ respectively.
In order to fix the gauge invariance of the term $\partial_{[\mu} \bar{\gamma}_{\nu]}^{-} \partial^{[\mu} {\gamma}^{-\nu]}$,
we have the Grassmann odd fields $\beta^{{\rm o}-}$ from ${\cal B}_{3,-}$ and $\bar{\beta}^{{\rm o}-}$ from ${\cal B}_{5,-}$, and the 
Grassmann even scalar fields $f$ from $b_0\tilde{c}_0\Phi_{0,-}$ and $\bar{f}$ from $b_0\tilde{c}_0\Phi_{5,-}$. 
The other fields appearing only in (\ref{eq:Sgf-2}) are auxiliary fields from ${\cal B}_{n,-}$ for $n=1,2,3$ and $\tilde{c}_0b_0 \Phi_{n,-}$ for $n=1,3,4,5$.

%%%%%%%%%%%%%%%%%%%%%%
\section*{Acknowledgements}
The work of M.K. is supported in part by the Grants-in-Aid for Scientific Research (No. 20340048) from the Japan Society for the Promotion of Science (JSPS).
%%%%%%%%%%%%%%%%%%
%%%%%%

%%%%%%%%%%%%%%%%%%%%%%%%%%%%%%%%%%%%%%%%%%%%%%

\end{document}